\documentclass[12pt]{article}
\usepackage{graphicx}
\usepackage{jheppub}
\title{RG domain wall for the N=1 minimal superconformal models}

\author[]{Gabriel Poghosyan \&}
\author[]{Hasmik Poghosyan}

\affiliation[]{Yerevan Physics Institute\\
Alikhanian Br. 2, 0036 Yerevan, Armenia}
\emailAdd{gabrielpoghos@gmail.com}
\emailAdd{hasmikpoghos@gmail.com}

\abstract{We specify Gaiotto's proposal for the
RG domain wall between some coset CFT models to
the case of two minimal N=1 SCFT models $SM_p$ and
$SM_{p-2}$  related by the RG flow initiated by the
top component of the Neveu-Schwarz superfield
$\Phi_{1,3}$ . We explicitly calculate the mixing
coefficients for several classes of fields and compare
the results with the already known in literature
results obtained through perturbative analysis.
Our results exactly match with both leading and next to
leading order perturbative calculations.
}

\begin{document}
\maketitle
\newcommand{\ie}{{\it i.e.\ }}
\def\bea{\begin{eqnarray}}
\def\eea{\end{eqnarray}}
\def\a{\alpha}
\def\b{\beta}
\def\g{\gamma}
\def\G{\Gamma}
\def\d{\delta}
\def\D{\Delta}
\def\e{\epsilon}
\def\z{\zeta}
\def\th{\theta}
\def\k{\kappa}
\def\l{\lambda}
\def\m{\mu}
\def\n{\nu}
\def\r{\rho}
\def\s{\sigma}
\def\t{\tau}
\def\f{\phi}

\section*{Introduction}

The existence of a RG flow between two CFT's suggests
that this theories could be connected by a non-trivial
interface which encodes the map from the UV observables
to the IR ones \cite{Fredenhagen:2005an, Brunner:2007ur}
In particular in \cite{Brunner:2007ur} such an interface
(RG domain wall) was constructed for the $N=2$ superconformal
models using matrix factorisation technique.

Later in \cite{Gaiotto:2012np} an algebraic construction
of a RG domain wall for the unitary minimal CFT models
was proposed and was shown that the results agree with
those of the leading order perturbative analysis performed by
A. Zamolodchikov in \cite{Zamolodchikov:1987ti}.

The leading order perturbative
calculation of the mixing coefficients for the wider class of
local fields including non-primary ones again is in an impressive
agreement with the RG domain wall approach \cite{Poghosyan:2013qta}.

Higher order perturbative calculations
\cite{Poghossian:2013fda, Konechny:2014opa} further
confirm the validity of this construction.

In the same paper \cite{Gaiotto:2012np} Gaiotto suggests
that a similar construction should be valid also for more
general coset CFT models. The $N=1$ minimal superconformal
CFT models \cite{Eichenherr:1985cx, Bershadsky:1985dq,
Friedan:1984rv}, which are the main subject of this paper,  are
among these cosets.

The Renormalisation Group (RG) flow between minimal
$N=1$ superconformal models
$SM_p$ and $SM_{p-2}$ initialised by the perturbation
with the top component of the Neveu-Schwarz superfield
$\Phi_{1,3}$ in leading order of the perturbation theory
has been investigated in \cite{Pogosian:1987zn} (see also
\cite{Kastor:1988ef, Crnkovic:1989gy}). 

Recently, extending the technique
developed in \cite{Poghossian:2013fda} for the minimal models to the
supersymmetric case, in \cite{Ahn:2014rua}
the analysis of this RG flow has been sharpened  even further 
by including also the next to leading order corrections.

In this paper we specialise Gaiotto's proposal to
the case of the minimal N=1 SCFT models. The method
we use is based directly on the current algebra construction
and, in this sense, is more general than the one originally
employed by Gaiotto for the case of minimal models. Namely
he heavily exploited  the fact that the product of successive
minimal models can be alternatively represented as a product
of $N=1$ superconformal and Ising models.
We explicitly calculate the mixing coefficients for several
classes of fields and compare the results with the perturbative
analysis of \cite{Pogosian:1987zn,Ahn:2014rua} finding a
complete agreement.

The paper is organised as follows:\\
Chapter \ref{SCFT revew} is a brief review of the 2d
$N=1$ superconformal filed theories.\\
Chapter \ref{Coset construction} is devoted to the
description of the coset construction of $N=1$ SCFT. Of course
everything here is well known; our purpose here is
to fix notations and list the relevant formulae in a form, most 
convenient for the further calculations.\\
In chapter \ref{General RG DW} we formulate Gaiotto's general 
proposal for a class of coset CFT models.\\
Chapter \ref{DW for SCFT} is the main part of our paper.
We explicitly calculate the mixing coefficients for the several
classes of local fields in the case of the supersymmetric RG flow 
discussed above using RG domain wall proposal. Then we compare 
this with the perturbation theory results available in the 
literature finding a complete agreement.

\section{N=1 superconformal filed theory}
\label{SCFT revew}
In any conformal filed theory the energy-momentum tensor
has two nonzero components: the holomorphic field $T(z)$ with
conformal dimension $(2,0)$ and its anti-holomorphic
counterpart $\bar{T}(\bar{z})$ with dimensions $(0,2)$.
In $N=1$ superconformal field theories one has in addition
superconformal currents $G(z)$ and $\bar{G}(\bar{z})$ with
dimensions $(3/2,0)$ and $(0,3/2)$ respectively.
These fields satisfy the OPE rules
\begin{eqnarray}
\label{TTOPE}
T(z)T(0)&=&\frac{c}{2z^4}+\frac{2T(0)}{z^2}+\frac{T^\prime(0)}{z}  + \cdots \\
\label{TGOPE}
T(z)G(0)&=&\frac{3G(0)}{2 z^{2}}+\frac{G^\prime(0)}{z} +\cdots\\
\label{GGOPE}
G(z)G(0)&=& \frac{2c}{3z^3}+\frac{2T(0)}{z}  +\cdots
\end{eqnarray}
The corresponding expressions for the anti-chiral fields look exactly the same. One should
 simply substitute $z$ by $\bar{z}$. Further on we'll mainly concentrate on the
   holomorphic part assuming similar expressions for anti-holomorphic
   quantities implicitly.
   We can expand $T(z)$  in Laurent series
\begin{equation}
T(z)=\sum \limits_{n=-\infty}^{+\infty} \frac{L_n}{z^{n+2}},
\end{equation}
where $L_n$'s are the Virasoro generators.

Due to the fermionic nature of the super current, there are two
distinct possibilities for its behavior under the rotation of its argument around
$0$ by  the angle $2 \pi$
\begin{align}
\label{NS monodromy}
G(e^{2\pi i }z)&=G(z)  \qquad \qquad \textit{Neveu - Schwarz sector (NS)} \\
\label{R monodromy}
G(e^{2\pi i }z)&=-G(z) \> \quad \qquad \textit{Ramond sector (R)}
\end{align}
The space of fields $\mathcal{A}$ of the superconformal theory decomposes
into a direct sum
\begin{equation}
\mathcal{A} = \{ NS\} \oplus \{R\}
\end{equation}
where the subspaces $\{ NS\}$ and $\{R\}$ consist of the Neveu-Shwarz
and the Ramond fields respectively. By definition, the monodromy of $S(z)$
around a Neveu-Schwarz field is trivial (the case of eq. (\ref{NS monodromy})) and
its monodromy around a Ramond field produces a minus sign (the case of eq.
(\ref{R monodromy})). Because of these two possibilities the
Laurent expansions for the super-current will be
\begin{eqnarray}
G(z) &=&\sum_{k \in Z+1/2} \frac{G_{k}}{z^{k+3/2}} \qquad \qquad
\textit{Neveu-Schwarz sector (NS)} \nonumber \\
G(z) &=&\sum_{k \in Z} \frac{G_{k}}{z^{k+3/2}} \qquad  \qquad \qquad
\textit{Ramond sector (R)}\nonumber
\end{eqnarray}
The OPE's (\ref{TTOPE}), (\ref{TGOPE}), (\ref{GGOPE}) are equivalent to the
Neveu-Schwarz-Ramond
algebra relations
\begin{eqnarray}
[L_n,L_m] &=& (n-m)L_{n+m} +\frac{c}{12}(n^3-n)
\delta_{n+m,0} \, ; \nonumber\\
\left[L_n, G_k\right] &=&\frac{1}{2}(n-2k)G_{n+k} \, ; \\
\{G_k,G_l\} &=& 2L_{k+l}+\frac{c}{3}\left(k^2-1/4\right)
\delta_{k+l,0} \, ;\nonumber
\end{eqnarray}
where $\{,\} $ denotes the anticommutator.
In this paper we'll deal with minimal super-conformal series denoted as $SM_p$ ($p=3,4,5\ldots $)
corresponding to the choice of the central charge
\begin{equation}
c_p=\frac{3}{2}\left(1-\frac{8}{p(p+2)}\right),
\label{SM central charge}
\end{equation}
The main distinctive mark of the minimal super-conformal theories is that they have finitely many  super primary fields.
These fields  are numerated by two integers   $n\in \{1,2,\cdots ,p-1$\}, $m\in \{1,2,\cdots ,p+1\}$ and will be denoted as $\phi_{n,m}$. It is assumed that   $\phi_{p-n,p+2-m}\equiv\phi_{n,m}$ , hens the number of super primaries  is equal to $[p^2/2]$ ([x] is the integer part of x).  $\phi_{p-1,p+1}\equiv\phi_{1,1}$ is the identity operator.  For even (odd) $n-m$
 the super-conformal classes $[\phi_{n,m}]$ form irreducible representations of the Neveu-Schwarz (Ramond) algebra. The fields $\phi_{n,m}$ have dimensions
\begin{equation}
h_{n,m}=\frac{((p+2)n-pm)^2-4}{8p(p+2)}+\frac{1}{32}(1-(-)^{n-m})
\label{Sminimal_dim}
\end{equation}

\section{Current algebra and the coset construction}
\label{Coset construction}
We will use the coset construction
\cite{Goddard:1984vk, Goddard:1986ee}
 of super-minimal models in terms of $\widehat{SU}(2)_k$ WZNW models
\cite{Knizhnik:1984nr, Zamolodchikov:1986bd}.

Remind that WZNW models are endowed with spin one holomorphic currents.
The OPE relations of these currents specified to the case of  $\widehat{SU}(2)_k$ read:
\begin{eqnarray}
\label{JOPE}
J^0(z)J^0(0)&=&\frac{k/2}{z^2}+reg \nonumber\\
J^0(z)J^\pm(0)&=&\pm\frac{J^\pm (0)}{z}+reg \\
J^+(z)J^-(0)&=&\frac{k}{z^2}+\frac{2J^0(0)}{z}+reg\nonumber
\end{eqnarray}
where $k$ is the level. The isotopic indices $\pm, 0$
convenient for the later use are related to the usual
Euclidean indices as:
\begin{eqnarray}
J^0&\equiv&J^3 \qquad \text{and}  \qquad  J^\pm \equiv J^1\pm iJ^2
\end{eqnarray}
The Laurent expansion of the currents reads
\begin{align}
J^a(z)=\sum\limits_{n \in Z}\frac{J^a_n}{z^{n+1}}
\end{align}
and the OPE rules (\ref{JOPE}) imply that the current
algebra generators are subject to the ${\text Ka\breve{c}-Moody}$ algebra commutation relations
\begin{eqnarray}
&&\left[J^{\pm}_n,J^{\pm}_m\right]= 0\nonumber\\
&&\left[J^+_n,J^-_m\right]= k n \delta_{n+m,0}+2J^0_{n+m}\nonumber\\
&&\left[J^0_n,J^{\pm}_m\right]=\pm J^{\pm}_{n+m} \\
&&\left[J^0_n,J^0_m\right]=\frac{k n}{2} \delta_{n+m,0}\nonumber
\label{Kac_Moody}
\end{eqnarray}
Notice that the subalgebra generated by $J^a_0$ is simply the
Lie algebra $su(2)$.\\
The energy momentum tensor can be expressed through the currents
with the help of the Sugawara consruction
 \begin{align}
 T(z)=\frac{1}{k+2} \left(J^0J^0+\frac{1}{2}J^+J^-
 +\frac{1}{2}J^-J^+\right)
\end{align}
As it is custom in CFT above and in what follows
we assume that any product of local fields taken at
coinciding points is regularised subtracting singular
parts of the respective OPE. The central charge of the Virasoro
algebra can be easily computed using (\ref{Kac_Moody}).
The result is:
\begin{eqnarray}
c_k=\frac{3k}{k+2}
\label{KM central charge}
\end{eqnarray}
The primary fields of the theory $\phi_{j,m}$ and
corresponding states $|j,m\rangle $ are labeled by the
spin of the representation $j=0,1/2,1,\ldots , k/2$
and its projection $m= -j,-j+1,\ldots ,j$. The
corresponding conformal dimensions are given by
\bea
h=\frac{j(j+1)}{k+2}
\label{WZW_dim}
\eea
The zero modes of the currents act on the states
$|j,m\rangle $ as \footnote{Note that a consistent
with eq. (\ref{current_0mode_action}) conjugation rule
for the primary fields would be
$\phi_{j,m}^\dagger =(-)^{j-m}\phi_{j,-m}$}
\begin{eqnarray}
J^{\pm}|j,m\rangle&=&\sqrt{j(j+1)-m(m\pm1)}|j,m\pm1\rangle
\nonumber\\
J^0|j,m\rangle&=&m|j,m\rangle
\label{current_0mode_action}
\end{eqnarray}
We'll need also the explicit form of the $su(2)$
WZNW modular matrices
\bea
S^{(k)}_{n,m} = \sqrt{\frac{2}{k+2}} \sin \frac{\pi n m}{k+2}
\label{su2_mod_matrix}
\eea

It is well known that the $N=1$ super-minimal models can be
represented as a coset \cite{Goddard:1984vk, Goddard:1986ee}
\begin{equation}
{\cal SM}_{k+2}=\frac{su(2)_k \times su(2)_2}{su(2)_{k+2}}\nonumber
\end{equation}
In particular the energy momentum tensor of ${\cal SM}_{k+2}$
is given by
\begin{equation}
T_{(su(2)_k\times su(2)_2)/su(2)_{k+2}}=T_{su(2)_k}+T_{su(2)_2}-T_{su(2)_{k+2}}
\end{equation}
Indeed the combination of the central charges
(\ref{KM central charge}) corresponding
to these three terms matches with the central charge of the
super-minimal models (\ref{SM central charge}).

The construction of the super-current $G$ is more subtle;
it involves the primary fields $\phi_{1,m}$ of the level
$k=2$ WZNW theory (we denote the currents of this theory as $K^a$ and 
summation over the index $a=\pm ,0$ is assumed):
\bea
G(z)=C_aJ^a(z)\phi_{1,-a}(z)+D_aK^a_{-1}\phi_{1,-a}(z)
\label{super_current_coset}
\eea
The coefficients $C_a$, $D_a$ can be fixed requiring that
the respective state be the highest weight state of the diagonal
current algebra $J+K$. In other words both $J^+_0+K^+_0$
and $J^+_1+K^+_1$ annihilate the state
\bea
C_aJ_{-1}^a|0\rangle|1,-a\rangle+D^a|0\rangle K^a_{-1}|1,-a \rangle .
\eea
Up to an overall constant $\kappa$ we get
\bea
\label{C_D_coefficients}
&&D_+=\frac{\kappa}{\sqrt{2}}\,;\hspace{1.4cm}  D_0 =\kappa \,; \hspace{1.4cm}   D_-=-\frac{\kappa}{\sqrt{2}}\nonumber \\
&&C_+=-\frac{3 \kappa\sqrt{2}}{k} \, ;\qquad C_0=-\frac{6\kappa}{k}\, ;
\qquad  C_-=\frac{3 \kappa\sqrt{2}}{k}
\eea
The value of $\kappa$ may be determined using the normalization
condition of the the super-current fixed by the OPE (\ref{GGOPE})
\bea
\kappa =\sqrt{\frac{(k+2)(k+4)}{(k+6)(5k+54)}}
\eea
but this wan't be of importance for our goals.

\section{Pertubative $\text{RG}$ flows and domain walls }
\label{General RG DW}
In a well known paper A. Zamolodchikov \cite{Zamolodchikov:1987ti}
has investigated the RG flow from minimal model
${\cal M}_p$  to  ${\cal M}_{p-1}$ initiated by the
relevant field $\phi_{1,3}$. Using leading order perturbation
theory valid for $p>>1$, for the several classes of local fields
he calculated the mixing coefficients specifying
the UV - IR map.

It was shown in \cite{Pogosian:1987zn} that a similar
RG trajectory connecting ${\cal N}=1$ super-minimal models
${\cal SM}_p$  to  ${\cal SM}_{p-2}$ exists. In this case the
RG flow is initiated by the top component of the Neveu-Schwartz
superfield $\Phi_{1,3}$. For us it will be important that
also in this case a detailed analysis of some classes of
fields has been carried out.

As it became clear later \cite{Crnkovic:1989gy,
Ravanini:1992fs}, above two examples are
just the first simplest cases of more general RG flows.
A wide class of CFT coset models
\begin{eqnarray}
 \mathcal{T}_{UV}&=&\frac{\hat{g}_l \times \hat{g}_m}{\hat{g}_{l+m}} \qquad m>l
 \label{GenUV}
 \end{eqnarray}
under perturbation by the relevant field $\phi=\phi^{Adj}_{1,1}$ \cite{Ravanini:1992fs} at the IR limit flow to the theories
\begin{eqnarray}
 \mathcal{T}_{IR}&=&\frac{\hat{g}_l \times \hat{g}_{m-l}}{\hat{g}_{m}}
\label{GenIR}
\end{eqnarray}

Recently in \cite{Gaiotto:2012np} Gaiotto constructed
a nontrivial conformal interface between successive
minimal CFT models and made a striking proposal that
this interface (RG domain wall) encodes the UV - IR map
resulting through the RG flow discussed above.
It was shown that the proposal agrees with
the leading order perturbative analysis of
\cite{Zamolodchikov:1987ti}.

Generalization of leading order calculations to
a wider class of local fields  \cite{Poghosyan:2013qta}
as well as next to leading order calculations
\cite{Poghossian:2013fda,Konechny:2014opa} further
confirm the validity of this construction.

Actually in \cite{Gaiotto:2012np} Gaiotto suggests
also a candidate for RG domain wall for the much more
general RG flow between (\ref{GenUV}) and (\ref{GenIR}).
Let us briefly recall the construction. Since a conformal
interface between two CFT models is equivalent to some
conformal boundary for the direct product of these theories
(folding trick), it is natural to consider the product theory
 $\mathcal{T}_{UV}\times \mathcal{T}_{IR}$
\begin{eqnarray}
\frac{\hat{g}_l \times \hat{g}_{m}}{\hat{g}_{m+l}}
\times \frac{\hat{g}_l \times \hat{g}_{m-l}}{\hat{g}_{m}}
\sim \frac{\hat{g}_{m-l}\times \hat{g}_l \times \hat{g}_l }{\hat{g}_{l+m}}
\end{eqnarray}
Notice the appearance of two identical factors $\hat{g}_l$
and one has a natural ${\mathbb{Z}}_2$ automorphism.
Essentially the proposal of Gaiotto boils down to the
statement that the boundary of the theory
\begin{eqnarray}
\mathcal{T}_B=\frac{\hat{g}_l \times \hat{g}_l \times \hat{g}_{m-l} }{\hat{g}_{l+m}}, \qquad m>l
\end{eqnarray}
acts as a ${\mathbb{Z}}_2$ twisting mirror. Explicitly the
RG boundary condition is the image of the ${\mathbb{Z}}_2$
twisted $\mathcal{T}_B$ brane
\begin{equation}
|\tilde{B} \rangle = \sum\limits_{s,t}\sqrt{S^{(m-l)}_{1,t}S^{(m+l)}_{1,s}}\sum\limits_d |t,d,d,s;\mathcal{B},Z_2\rangle \rangle ,
\label{Gen_boundary}
\end{equation}
where the indices $t$, $d$, $s$ refer to the representations
of $\hat{g}_{m-l}$, $\hat{g}_l$, $\hat{g}_{l+m}$ respectively
and $S^{(k)}_{1,r}$ are the modular matrices of the
$\hat{g}_{k}$ WZNW model.

In what follows we will examine in details the case of
RG flow between ${\cal N}=1$ super-minimal models. The method
we apply directly explores the current algebra representation
in contrary to the analysis in \cite{Gaiotto:2012np} where
a specific representation applicable only for the unitary minimal
series was used.

\section{$\text{RG}$ domane walls for super minimal models}
\label{DW for SCFT}
In the case of the ${\cal N}=1$ super-minimal models
one should consider
\begin{equation}
\frac{\widehat{su}(2)_k\times \widehat{su}(2)_2}
{\widehat{su}(2)_{k+2}}\times\frac{\widehat{su}(2)_{k-2}
\times \widehat{su}(2)_2}{\widehat{su}(2)_k}\sim 
\frac{\widehat{su}(2)_{k-2}\times \widehat{su}(2)_2 
\times \widehat{su}(2)_2}{\widehat{su}(2)_{k+2}}
\end{equation}
where the first coset on lhs corresponds to the UV
super conformal model ${\cal SM}_{k+2}$ and the second
one to the IR theory ${\cal SM}_{k}$.
We denote by $K(z)$ and $\widetilde{K}(z)$ the WZNW currents
of $\widehat{su}(2)_2$ entering in the cosets of the IR and UV
theories respectively. The current of $\widehat{su}(2)_{k-2}$ WZNW
theory will be denoted as $J(z)$. Using  Sugawara
construction for the energy-momentum tensor of the IR theory
we get
\bea
T_{ir}(z)=\frac{2}{2k+k^2}J(z)J(z)
-\frac{2}{2+k}J(z)K(z)+\frac{k-2}{4(k+2)}K(z)K(z)
\label{T_ir}
\eea
Similarly the energy-momentum tensor for the UV
theory is equal to
\bea
T_{uv}(z)=\frac{2}{(2+k)(4+k)}J(z)J(z)
+\frac{2}{(2+k)(4+k)}K(z)K(z)\nonumber \\
-\frac{2}{4+k}K(z)\widetilde{K}(z)+\frac{k}{4(k+4)}
\widetilde{K}(z)\widetilde{K}(z)\nonumber\\
+\frac{4}{(2+k)(4+k)}J(z)K(z)-\frac{2}{4+k}J(z)\widetilde{K}(z)
\eea
In order to get the one-point functions of the
theory ${\cal SM}_{k+2}\times {\cal SM}_{k}$ in the
presence of RG boundary, one needs explicit expressions
of the states corresponding to fields $\phi^{IR}\phi^{UV}$
in terms of the states of the coset theory
\bea
{\cal T}_B=\frac{\widehat{su}(2)_{k-2}\times \widehat{su}(2)_2 
\times \widehat{su}(2)_2}
{\widehat{su}(2)_{k+2}}
\eea
Let us denote the highest weight representation spaces of the current
algebras $J(z)$, $K(z)$ and ${\widetilde K}(z)$
as $V_{j}^{(J)}$, $V_k^{(K)} $ and $V_{\widetilde{k}}^{(\widetilde{K})}$
respectively (the lower indices specify the spins of the highest
weight states). It is convenient to fix a unique representative
of a state of coset ${\cal T}_B$ in space
$V_{j}^{(J)}\otimes V_k^{(K)} \otimes V_{\widetilde{k}}^{(\widetilde{K})}$
requiring that the state under consideration be a highest weight
state of the diagonal current $J+K+\widetilde{K}$. The simplest case
to analyse are the states corresponding to
$\phi_{n,n}^{IR}\phi_{n,n}^{UV}$. Since
\bea
&&h^{ir}_{n,n}=\frac{n^2-1}{4k}-\frac{n^2-1}{4(k+2)}\nonumber \\
&&h^{uv}_{n,n}=\frac{n^2-1}{4(k+2)}-\frac{n^2-1}{4(k+4)},\nonumber
\eea
the total dimension of the product field is
\bea
\label{nn_nn_state}
h^{ir}_{n,n}+h^{uv}_{n,n}=\frac{n^2-1}{4k}-\frac{n^2-1}{4(k+4)}
\eea
so that the corresponding state is readily identified with
($|j,m\rangle$ denotes a primary state of spin $j$ and
projection $m$)
\bea
|\frac{n-1}{2},\frac{n-1}{2}\rangle |0,0\rangle|0,0\rangle
\in V_{\frac{n-1}{2}}^{(J)}\otimes V_0^{(K)} \otimes V_{0}^{(\widetilde{K})}
\eea
Indeed, this is a spin $\frac{n-1}{2}$ highest weight state of 
the combined current
$J+K+\widetilde{K}$ and its ${\cal T}_B$ dimension
\bea
h_{\frac{n-1}{2}}^{(J)}+h_{0}^{(K)}+h_{0}^{({\widetilde K})}
-h_{\frac{n-1}{2}}^{(J+K+{\widetilde K})}\nonumber
\eea
coincides with (\ref{nn_nn_state}). Notice that $\mathbb{Z}_2$
action (i.e. permutation of the second and third factors) on
this state is trivial. Thus the overlap of this state with
its $\mathbb{Z}_2$ image is equal to 1 and from (\ref{Gen_boundary})
\bea
\langle \phi^{IR}_{n,n} \phi^{UV}_{n,n} | RG \rangle =  \frac{\sqrt{S^{(k-2)}_{1,n} S^{(k+2)}_{1,n}}}{S^{(k)}_{1,n}}
\eea
For large $k$ and for $n\sim O(1)$ this gives $1+3/k^2+O(1/k^3)$.
We conclude that up to $1/k^2$ terms, the fields $\phi^{UV}_{n,n}$
flow to $\phi^{IR}_{n,n}$ without mixing with other fields,
in complete agreement with both leading order \cite{Pogosian:1987zn}
and next to leading order \cite{Ahn:2014rua} perturbative calculations.

Next let us examine the more interesting case of Ramond fields
$\phi^{UV}_{n,n\pm 1}$ which are expected to flow to certain
combinations of the fields $\phi^{IR}_{n\pm 1,n}$
\cite{Pogosian:1987zn}.

Consider the state corresponding to
$\phi_{n-1,n}^{ir}\phi_{n,n-1}^{uv}$. From
(\ref{Sminimal_dim})we get
\begin{gather}
\label{hirnm1n}
h^{ir}_{n-1,n}=\frac{3}{16}+\frac{(n-1)^2-1}{4k}
-\frac{n^2-1}{4(k+2)} \\
\label{huvnnm1}
h^{uv}_{n,n-1}=\frac{3}{16}-\frac{(n-1)^2-1}{4(k+4)}+\frac{n^2-1}{4(k+2)}
\end{gather}
Hence the conformal dimension of this product field will be
\begin{gather}
h^{ir}_{n-1,n}+h^{uv}_{n,n-1}=\frac{3}{8}
+\frac{(n-1)^2-1}{4k}-\frac{(n-1)^2-1}{4(k+4)}
\label{dim_nm1nm1}
\end{gather}

There are three primaries in $su(2)_2$ WZNW theory with $j=0,1,2$
representations and conformal dimensions  $0$,$\frac{3}{16}$
and $\frac{1}{2})$ respectively. So, to
get the right dimension one should choose a combination
of states $|\frac{n}{2}-1,m\rangle 
|\frac{1}{2},\alpha\rangle|\frac{1}{2},\beta\rangle$.
In addition this combination must be the spin $\frac{n}{2}-1$ highest weight
state of $J+K+\widetilde{K}$ (to match with the last, negative
term of (\ref{dim_nm1nm1}) ). Thus we are lead to
\begin{gather}
C_{\alpha \beta}|\frac{n}{2}-1,\frac{n}{2}-1-\alpha-\beta\rangle |\frac{1}{2},\alpha\rangle|\frac{1}{2},\beta\rangle ,
\label{nm1nm1state}
\end{gather}
where a summation over the indices $\a, \b =\pm 1/2$ is assumed.
The highest weight condition that the operator
$J_0^++K^+_0+\widetilde{K}_0$ annihilates this state,
implies
\bea
\sqrt{n-2}C_{++}+C_{-+}+C_{+-} =0.\nonumber
\eea
A further constraint
\bea
C_{++}-\sqrt{n-2}C_{-+}=0\nonumber
\eea
one obtains imposing the condition
that this state should be an eigenstate of the Virasoro
operator $L_0^{IR}$ constructed from the energy-momentum tensor
$T_{ir}$ (\ref{T_ir}) with egenvalue $h_{n,n-1}^{ir}$
(\ref{hirnm1n}). Thus we get
\bea
C_{++}=\sqrt{n-2}\,C_{-+}\, ;\qquad C_{+-}=-(n-1)\,C_{-+}\nonumber
\eea
(of course, the undefined overall multiplier could be fixed
from the normalization condition). Taking (normalized) scalar
product of the state (\ref{nm1nm1state}) with its $\mathbb{Z}_2$
image we find
\bea
\langle\phi^{ir}_{n-1,n}\phi^{uv}_{n,n-1}|RG\rangle=
-\frac{1}{n-1}\frac{\sqrt{S^{(k-2)}_{1,n-1}
S^{(k+2)}_{1,n-1}}}{S^{k}_{1,n}}
\eea
Consideration of the product $\phi_{n+1,n}^{ir}\phi_{n,n+1}^{uv}$
fields is quite similar and leads to the state
\bea
C_{\alpha \beta}|\frac{n}{2},\frac{n}{2}-\alpha-\beta\rangle |\frac{1}{2},\alpha\rangle|\frac{1}{2},\beta\rangle\nonumber
\eea
with the coefficients
\bea
C_{+-}=0 \,;\qquad
C_{++}=-\frac{1}{\sqrt{n}}C_{-+}\nonumber
\eea
So, in this case
\bea
\langle\phi^{ir}_{n+1,n}\phi^{uv}_{n,n+1}|RG\rangle
=\frac{1}{n+1}\frac{\sqrt{S^{(k-2)}_{1,n+1}
S^{(k+2)}_{1,n+1}}}{S^{k}_{1,n}}
\eea
Constructing the states corresponding to
$\phi_{n-1,n}^{ir}\phi_{n,n+1}^{uv}$ and
$\phi_{n+1,n}^{ir}\phi_{n,n-1}^{uv}$ is even simpler and
one easily gets $|\frac{n}{2}-1,\frac{n}{2}-1\rangle |\frac{1}{2},\frac{1}{2}\rangle|\frac{1}{2},\frac{1}{2}\rangle
$
and
$
|\frac{n}{2},\frac{n}{2}\rangle |\frac{1}{2},-\frac{1}{2}
\rangle|\frac{1}{2},-\frac{1}{2}\rangle
$
respectively. In both cases the $\mathbb{Z}_2$
action is trivial, hence
\begin{gather}
\langle\phi^{ir}_{n-1,n}\phi^{uv}_{n,n+1}|RG\rangle
=\frac{\sqrt{S^{(k-2)}_{1,n-1}S^{(k+2)}_{1,n+1}}}{S^{k}_{1,n}}\\
\langle\phi^{ir}_{n+1,n}\phi^{uv}_{n,n-1}|RG\rangle
=\frac{\sqrt{S^{(k-2)}_{1,n+1}S^{(k+2)}_{1,n-1}}}{S^{k}_{1,n}}
\end{gather}
In the large $k$ limit we get
\begin{eqnarray}
\langle\phi^{ir}_{n+1,n}\phi^{uv}_{n,n+1}|RG\rangle
&=&\frac{1}{n}+O(1/k^2)\\
\langle\phi^{ir}_{n+1,n}\phi^{uv}_{n,n-1}|RG\rangle
&=&\frac{\sqrt{n^2-1}}{n}+O(1/k^2)\\
\langle\phi^{ir}_{n-1,n}\phi^{uv}_{n,n+1}|RG\rangle
&=&\frac{\sqrt{n^2-1}}{n}+O(1/k^2)\\
\langle\phi^{ir}_{n-1,n}\phi^{uv}_{n,n-1}|RG\rangle
&=&-\frac{1}{n}+O(1/k^2)
\end{eqnarray}
in complete agreement with the second order perturbation
theory results \cite{Ahn:2014rua}.

We have analysed also the more complicated case of mixing
of the primary Neveu-Schwartz superfields $\Phi_{n,n\pm 2}$
and the descendant superfield
$\mathbf{D}\bar{ \mathbf{D}}\Phi_{n,n}$
(here $\mathbf{D}$ and $\bar{ \mathbf{D}}$ are the
super-derivatives). The details of calculations are presented
in the appendix. Here are the final results:
\bea
&&\langle\psi^{ir}_{n+2,n}\psi^{uv}_{n,n+2}|RG\rangle
=\frac{2}{(n+1)(n+2)}
\frac{\sqrt{S^{(k-2)}_{1,n+2}S^{(k+2)}_{1,n+2}}}{S^{(k)}_{1,n}}\\
&&\langle \phi^{ir}_{n+2,n}G^{uv}_{-\frac{1}{2}}\phi^{uv}_{n,n}|RG\rangle
=\frac{2}{n+1}\frac{\sqrt{S^{(k-2)}_{1,n+2}
S^{(k+2)}_{1,n}}}{S^{(k)}_{1,n}}\\
&&\langle\psi^{ir}_{n+2,n}\psi^{uv}_{n,n-2}|RG\rangle
=\frac{\sqrt{S^{(k-2)}_{1,n+2}S^{(k+2)}_{1,n-2}}}{S^{(k)}_{1,n}}\\
&&\langle
G^{ir}_{-\frac{1}{2}}\phi^{ir}_{n,n}\phi^{uv}_{n,n+2}|RG\rangle
=\frac{2}{n+1}\frac{\sqrt{S^{(k-2)}_{1,n}
S^{(k+2)}_{1,n+2}}}{S^{(k)}_{1,n}}\\
&&\langle G_{-\frac{1}{2}}^{ir}\phi_{n,n}^{ir}
G_{-\frac{1}{2}}^{uv}\phi_{n,n}^{uv}|RG\rangle
=\frac{n^2-5}{n^2-1}\,\frac{\sqrt{S^{(k-2)}_{1,n}
S^{(k+2)}_{1,n}}}{S^{(k)}_{1,n}}
\eea
\bea
&&\langle G^{ir}_{-\frac{1}{2}}\phi^{ir}_{n,n}\phi^{uv}_{n,n-2}|RG\rangle
=-\frac{2}{n-1}\frac{\sqrt{S^{(k-2)}_{1,n}S^{(k+2)}_{1,n-2}}}{S^{(k)}_{1,n}}\\
&&\langle\psi^{ir}_{n-2,n}\psi^{uv}_{n,n+2}|RG\rangle
=\frac{\sqrt{S^{(k-2)}_{1,n-2}S^{(k+2)}_{1,n+2}}}{S^{(k)}_{1,n}}\\
&&\langle \phi^{ir}_{n-2,n}G^{uv}_{-\frac{1}{2}}\phi^{uv}_{n,n}|RG\rangle
=-\frac{2}{n-1}\frac{\sqrt{S^{(k-2)}_{1,n-2}
S^{(k+2)}_{1,n}}}{S^{(k)}_{1,n}}\\
&&\langle\phi^{ir}_{n-2,n}\phi^{uv}_{n,n-2}
|RG\rangle=\frac{2}{(n-1)(n-2)}
\frac{\sqrt{S^{(k-2)}_{1,n-2}S^{(k+2)}_{1,n-2}}}{S^{k}_{1,n}}
\end{eqnarray}
At the large $k$ limit we get
\begin{eqnarray}
&&\langle\psi^{ir}_{n+2,n}\psi^{uv}_{n,n+2}|RG\rangle
=\frac{2}{n(n+1)}+O(1/k^2)\\
&&\langle
\phi^{ir}_{n+2,n}G^{uv}_{-\frac{1}{2}}\phi^{uv}_{n,n}|RG\rangle
=\frac{2}{n+1}\sqrt{\frac{n+2}{n}}+O(1/k^2)\\
&&\langle\psi^{ir}_{n+2,n}\psi^{uv}_{n,n-2}|RG\rangle
=\frac{\sqrt{n^2-4}}{n}+O(1/k^2)\\
&&\langle
G^{ir}_{-\frac{1}{2}}\phi^{ir}_{n,n}\phi^{uv}_{n,n+2}|RG\rangle
=\frac{2}{n+1}\sqrt{\frac{n+2}{n}}+O(1/k^2)\\
&&\langle G_{-\frac{1}{2}}^{ir}\phi_{n,n}^{ir}
G_{-\frac{1}{2}}^{uv}\phi_{n,n}^{uv}|RG\rangle
=\frac{n^2-5}{n^2-1}+O(1/k^2)\\
&&\langle
G^{ir}_{-\frac{1}{2}}\phi^{ir}_{n,n}\phi^{uv}_{n,n-2}|RG\rangle
=-\frac{2}{n-1}\sqrt{\frac{n-2}{n}}+O(1/k^2)\\
&&\langle\psi^{ir}_{n-2,n}\psi^{uv}_{n,n+2}|RG\rangle
=\frac{\sqrt{n^2-4}}{n}+O(1/k^2)\\
&&\langle
\phi^{ir}_{n-2,n}G^{uv}_{-\frac{1}{2}}\phi^{uv}_{n,n}|RG\rangle
=-\frac{2}{n-1}\sqrt{\frac{n-2}{n}}+O(1/k^2)\\
&&\langle\phi^{ir}_{n-2,n}\phi^{uv}_{n,n-2}|RG\rangle
=\frac{2}{n(n-1)}+O(1/k^2)
\end{eqnarray}
Again, the results are in complete agreement with the next to leading
order perturbative calculations of \cite{Ahn:2014rua}

It is interesting to note that, though the mixing coefficients
computed here in the large $k$ limit coincide with
the respective cases of the $\phi_{1,3}$ perturbed
minimal models, the exact $k$ dependence in supersymmetric
case enters solely through the modular matrices, in contrary to the quite complicated $k$ dependence of the non supersymmetric case.
\section*{Acknowledgements}
We thank Ruben Manvelyan, Rubik Poghossian, and
Gor Sarkissian for
introducing us into this field of research.
The work of Gabriel Poghosyan was partially supported
by the Armenian SCS grant 13-1C232.
The work of Hasmik Poghosyan was partially supported
by the Armenian SCS grant 13-1C132 and by the
Armenian-Russian SCS grant-2013.
\begin{appendix}
\section{Mixing of the fields $\Phi_{n,n\pm 2}$
and the descendant
$\mathbf{D}\bar{ \mathbf{D}}\Phi_{n,n}$}
\label{A}
Let us start with the product field
$\phi_{n-2,n}^{ir}\phi_{n,n-2}^{uv}$. The corresponding
dimensions are
\begin{align}
\label{hirnnm2}
h^{ir}_{n-2,n}&=\frac{1}{2}+\frac{(n-2)^2-1}{4k}-\frac{n^2-1}{4(k+2)} \\
h^{uv}_{n,n-2}&=\frac{1}{2}-\frac{(n-2)^2-1}{4(4+k)}+\frac{n^2-1}{4(k+2)}
\end{align}
hence
\begin{equation}
h^{ir}_{n-2,n}+h^{uv}_{n,n-2}=1+\frac{(n-2)^2-1}{4k}-\frac{(n-2)^2-1}{4(4+k)}
\end{equation}
A careful examination shows that the required state should
be chosen among the combinations
\begin{gather}
\sum_{\alpha,\beta\in \{-1,0,1\}}C_{\alpha,\beta}|\frac{n-3}{2},\frac{n-3}{2}
-\alpha-\beta\rangle|1,\alpha\rangle|1,\beta\rangle
\label{state_nm2nnnm2}
\end{gather}
Indeed the other candidates such as
$
J^a_{-1}|\frac{n-3}{2},\frac{n-3}{2}-a\rangle|0\rangle|0\rangle\nonumber $,
$K^a_{-1}|\frac{n-3}{2},\frac{n-3}{2}-a\rangle|0\rangle|0\rangle$ or
$\widetilde{K}^{\a}_{-1}|\frac{n-3}{2},\frac{n-3}{2}-a\rangle|0\rangle|0\rangle$ though have a correct total dimension, can not be combined to
get the required IR dimension (\ref{hirnnm2}). This can be easily
seen by examining the zero mode of the IR current
\begin{equation}
T^{ir}=\frac{1}{k}J^2-\frac{1}{k+2}(J+K)^2+\frac{1}{4}K^2
\label{T_ir_n}
\end{equation}
The only way to get the term $1/2$ of (\ref{hirnnm2})
is to choose $j=1$ representation of the current $K$
(see the last term of (\ref{T_ir_n})).

To get correct IR dimension one should impose the
condition that the zero mode of $(J+K)^2$ on the state
(\ref{state_nm2nnnm2}) must acquire the eigenvalue
$\frac{n-1}{2}\frac{n+1}{2}$. Together with our usual
requirement of being a highest weight state of the
$J+K+\widetilde{K}$ algebra this fixes the coefficients
up to an overall multiplier
\begin{align}
C_{+0}&=\sqrt{\frac{n-3}{2}}C_{00}\,;\qquad & C_{++}&=-\sqrt{\frac{n-3}{2}}\frac{\sqrt{n-4}}{n-2}C_{00}
\nonumber\\
C_{+-}&=\frac{1-n}{2}C_{00}\, ; \qquad & C_{0+}&=-\frac{2}{n-2}
\sqrt{\frac{n-3}{2}}C_{00}\nonumber \\
C_{-+}&=-\frac{1}{n-2}C_{00}\, ; \qquad & C_{-0}&=C_{0-}
=C_{--}=0\nonumber
\end{align}
This leads to the one point function
\begin{gather}
\langle\phi^{ir}_{n-2,n}\phi^{uv}_{n,n-2}|RG\rangle
=\frac{2}{(n-1)(n-2)}\frac{\sqrt{S^{(k-2)}_{1,n-2}
S^{(k+2)}_{1,n-2}}}{S^{k}_{1,n}}
\end{gather}
In the same way we construct the state corresponding 
to $\phi_{n+2,n}^{ir}\phi_{n,n+2}^{uv}$
\begin{align}
C_{\alpha \beta}|\frac{n+1}{2},\frac{n+1}{2}-\alpha-\beta\rangle
|1,\alpha\rangle|1,\beta\rangle\nonumber
\end{align}
where
\begin{gather}
C_{++}=-\frac{1}{\sqrt{n}}\,C_{00}, \quad C_{-+}
=-\sqrt{\frac{n+1}{2}}\,C_{00},\quad
\quad C_{0+}=C_{00}
\end{gather}
(all other $C_{\alpha \beta}$ vanish) and
\begin{align}
\langle\psi^{ir}_{n+2,n}\psi^{uv}_{n,n+2}|RG\rangle
=\frac{2}{(n+1)(n+2)}\frac{\sqrt{S^{(k-2)}_{1,n+2}
S^{(k+2)}_{1,n+2}}}{S^{(k)}_{1,n}}
\end{align}
The state corresponding to $\psi^{ir}_{n+2,n}\psi^{uv}_{n,n-2}$
is simply
$|\frac{n+1}{2},\frac{n+1}{2}\rangle|1,-1\rangle|1,-1\rangle$
and
\bea
\langle\psi^{ir}_{n+2,n}\psi^{uv}_{n,n-2}|RG\rangle
=\frac{\sqrt{S^{(k-2)}_{1,n+2}S^{(k+2)}_{1,n-2}}}{S^{(k)}_{1,n}}
\eea
Similarly for $\psi^{ir}_{n-2,n}\psi^{uv}_{n,n+2}$ the state
is
$
|\frac{n-3}{2},\frac{n-3}{2}\rangle|1,1\rangle|1,1\rangle$
and
\bea
\langle\psi^{ir}_{n-2,n}\psi^{uv}_{n,n+2}|RG\rangle
=\frac{\sqrt{S^{(k-2)}_{1,n-2}S^{(k+2)}_{1,n+2}}}{S^{(k)}_{1,n}}
\eea

Let us now consider states corresponding to the descendant field $G^{ir}_{-1/2}\psi^{ir}_{n,n}\psi^{uv}_{n,n+2}$.

Partial dimensions of the field $\phi^{ir}_{n,n}\phi^{uv}_{n,n+2}$
are
\bea
&&h^{ir}_{n,n}= \frac{n^2-1}{4k}-\frac{n^2-1}{4(k+2)}\nonumber\\
&&h^{uv}_{n,n+2}=\frac{1}{2}+\frac{n^2-1}{4(k+2)}
-\frac{(n+2)^2-1}{4(k+4)}\nonumber\\
&&h^{ir}_{n,n}+h^{uv}_{n,n+2}=
\frac{1}{2}+\frac{n^2-1}{4k}-\frac{(n+2)^2-1}{4(k+4)}\nonumber
\eea
Evidently the correct representative of the
respective state is
\begin{align}
|\frac{n-1}{2},\frac{n-1}{2}\rangle|0\rangle|1,1\rangle .
\end{align}
Using the expression (\ref{super_current_coset}) its is straightforward
to find the result of the action of the super-current mode
$G_{-1/2}^{ir}$ on this state:
\begin{align}
G^{ir}_{-\frac{1}{2}}|\frac{n-1}{2},\frac{n-1}{2}
\rangle|0\rangle|1,1\rangle =C_aJ^a_0|\frac{n-1}{2},\frac{n-1}{2}
\rangle|1,-a\rangle|1,1\rangle
\nonumber \\
+D_aK^a_0|\frac{n-1}{2},\frac{n-1}{2}
\rangle|1,-a\rangle|1,1\rangle
\end{align}
where the coefficients $C_a$, $D_a$ are given by
(\ref{C_D_coefficients}) (one should replace $k$ by $k-2$).
The final result is:
\bea
G^{ir}_{-\frac{1}{2}}|\frac{n-1}{2},\frac{n-1}{2}
\rangle|0\rangle|1,1\rangle=
-\frac{3(n-1)}{k-2}|\frac{n-1}{2},\frac{n-1}{2}
\rangle|1,0\rangle|1,1\rangle \nonumber \\
+\frac{6}{k-2}\sqrt{\frac{n-1}{2}}
|\frac{n-1}{2},\frac{n-3}{2}\rangle|1,1\rangle|1,1\rangle
\eea
Thus for the one-point function we get
\bea
\langle G^{ir}_{-\frac{1}{2}}\phi^{ir}_{n,n}\phi^{uv}_{n,n+2}
|RG\rangle=\frac{2}{n+1}\frac{\sqrt{S^{(k-2)}_{1,n}
S^{(k+2)}_{1,n+2}}}{S^{(k)}_{1,n}}
\eea
Consideration of the remaining cases do not involve
new ingredients and we will simply list the results.

\begin{itemize}
\item{The state corresponding to $\phi^{ir}_{n,n}\phi^{uv}_{n,n-2}$ is:
\begin{eqnarray}
-\frac{1}{\sqrt{n-2}}|\frac{n-1}{2}
,\frac{n-5}{2}\rangle|0\rangle|1,1\rangle
+|\frac{n-1}{2},\frac{n-3}{2}\rangle |0\rangle |1,0\rangle
\nonumber \\
-\sqrt{\frac{n-1}{2}}|\frac{n-1}{2},
\frac{n-1}{2}\rangle|0\rangle|1,-1\rangle\nonumber
\end{eqnarray}
}
\end{itemize}
The result of $G^{ir}_{-\frac{1}{2}}$ action on this state looks 
ugly:
\begin{eqnarray}
|\frac{n-1}{2},\frac{n-3}{2}\rangle
|1,-1\rangle|1,1\rangle
+\frac{n-5}{2\sqrt{n-2}}|\frac{n-1}{2},\frac{n-5}{2}\rangle
|1,0\rangle|1,1\rangle
\nonumber \\
-\sqrt{\frac{3n-9}{2n-4}}
|\frac{n-1}{2},\frac{n-7}{2}\rangle|1,1\rangle|1,1\rangle
-\sqrt{\frac{n-1}{2}}
|\frac{n-1}{2},\frac{n-1}{2}\rangle|1,-1\rangle|1,0\rangle
\nonumber \\
-\frac{n-3}{2}|\frac{n-1}{2},\frac{n-3}{2}\rangle
|1,0\rangle|1,0\rangle
+\sqrt{n-2}|\frac{n-1}{2},\frac{n-5}{2}\rangle
|1,1\rangle|1,0\rangle
\nonumber \\
+\Big(\frac{n-1}{2}\Big)^{\frac{3}{2}}
|\frac{n-1}{2},\frac{n-1}{2}\rangle|1,0\rangle|1,-1\rangle
-\frac{n-1}{2}|\frac{n-1}{2},\frac{n-3}{2}\rangle
|1,1\rangle|1,-1\rangle\nonumber
\end{eqnarray}
multiplied by an overall factor $\frac{6}{k-2}$. 
The corresponding one-point function simply is:
\begin{equation}
\langle G^{ir}_{-\frac{1}{2}}\phi^{ir}_{n,n}\phi^{uv}_{n,n-2}
|RG\rangle=-\frac{2}{n-1}\frac{\sqrt{S^{(k-2)}_{1,n}
S^{(k+2)}_{1,n-2}}}{S^{(k)}_{1,n}}
\end{equation}
\begin{itemize}
\item{In the $\phi^{ir}_{n-2,n}\phi^{uv}_{n,n}$ case the
corresponding state is
\begin{equation}
|\frac{n-3}{2},\frac{n-3}{2}\rangle|1,1\rangle|0\rangle
\end{equation}
Now we must act on this state by the operator $G^{uv}_{-1/2}$
\bea
G^{uv}_{-1/2}|\frac{n-3}{2},\frac{n-3}{2}\rangle|1,1\rangle|0\rangle
=\left(C_a(K^a_{0}+J^a_{0})+D_a\widetilde{K}^{a}_{0}\right)
|\frac{n-3}{2},\frac{n-3}{2}\rangle|1,-a\rangle|0\rangle\nonumber\\
=-\frac{3(n-1)}{k}|\frac{n-3}{2},\frac{n-3}{2}\rangle
|1,1\rangle|1,0\rangle
+\frac{6}{k}|\frac{n-3}{2},\frac{n-3}{2}\rangle
|1,0\rangle|1,1\rangle\nonumber \\
+\frac{6}{k}\sqrt{\frac{n-3}{2}}
|\frac{n-3}{2},\frac{n-5}{2}\rangle|1,1\rangle|1,1\rangle\nonumber
\eea
The one point function:
\bea
\langle \phi^{ir}_{n-2,n}G^{uv}_{-\frac{1}{2}}\phi^{uv}_{n,n}
|RG\rangle=-\frac{2}{n-1}
\frac{\sqrt{S^{(k-2)}_{1,n-2}S^{(k+2)}_{1,n}}}{S^{(k)}_{1,n}}
\eea
}
\item{The state corresponding to the field
$\phi^{ir}_{n+2,n}\phi^{uv}_{n,n}$ is
\begin{eqnarray}
-\frac{1}{\sqrt{n}}|\frac{n+1}{2}
\frac{n-3}{2}\rangle|1,1\rangle|0\rangle
+|\frac{n+1}{2},\frac{n-1}{2}\rangle|1,0\rangle|0\rangle
\nonumber \\
-\sqrt{\frac{n+1}{2}}|\frac{n+1}{2},\frac{n+1}{2}\rangle
|1,-1\rangle|0\rangle
\end{eqnarray}
Acting by $G^{uv}_{-1/2}$ on this state we get
\bea
\frac{n-1}{2\sqrt{n}}|\frac{n+1}{2},\frac{n-3}{2}\rangle
|1,1\rangle|1,0\rangle +\sqrt{\frac{n+1}{2}}(\frac{n-1}{2})
|\frac{n+1}{2},\frac{n+1}{2}\rangle|1,-1\rangle|1,0\rangle
\nonumber \\
-\sqrt{\frac{3n-3}{2n}}|\frac{n+1}{2},\frac{n-5}{2}\rangle
|1,1\rangle|1,1\rangle
+\frac{n-1}{\sqrt{n}}|\frac{n+1}{2},\frac{n-3}{2}\rangle
|1,0\rangle|1,1\rangle
\nonumber \\
-\frac{n-1}{2}|\frac{n+1}{2},\frac{n-1}{2}\rangle
|1,0\rangle|1,0\rangle
-\frac{n-1}{2}|\frac{n+1}{2},\frac{n-1}{2}\rangle
|1,-1\rangle|1,1\rangle
\nonumber
\eea
multiplied by $\frac{6}{k}$.
The result for one-point function:
\begin{gather}
\langle \phi^{ir}_{n+2,n}G^{uv}_{-\frac{1}{2}}\phi^{uv}_{n,n}
|RG\rangle=\frac{2}{n+1}
\frac{\sqrt{S^{(k-2)}_{1,n+2}S^{(k+2)}_{1,n}}}{S^{(k)}_{1,n}}
\end{gather}
}
\item{Finally, the state corresponding to the field
$G_{-\frac{1}{2}}^{ir}\phi^{ir}_{n,n}
G_{-\frac{1}{2}}^{uv}\phi^{uv}_{n,n}$ is
\begin{equation}
(C_aJ^a_0+D_aK^a_0)(C_b(K^b_0+J^b_0)+D_b
\widetilde{K}^b_0)|\frac{n-1}{2},\frac{n-1}{2}\rangle
|1,-a\rangle|1,-b\rangle
\end{equation}
which after some algebra becomes
\begin{eqnarray}
\Big(\frac{n-1}{2}\Big)^2
|\frac{n-1}{2},\frac{n-1}{2}\rangle|1,0\rangle
|1,0\rangle
-\sqrt{\frac{n-1}{2}}\frac{n-1}{2}
|\frac{n-1}{2},\frac{n-3}{2}\rangle|1,0\rangle|1,1\rangle\nonumber\\
-\frac{n-1}{2}
|\frac{n-1}{2},\frac{n-1}{2}\rangle|1,1\rangle|1,-1\rangle
-\sqrt{\frac{n-1}{2}}\frac{n-3}{2}
|\frac{n-1}{2},\frac{n-3}{2}\rangle|1,1\rangle|1,0\rangle\nonumber\\
+\sqrt{\frac{n-1}{2}}\sqrt{n-2}
|\frac{n-1}{2},\frac{n-5}{2}\rangle|1,1\rangle|1,1\rangle\nonumber
\end{eqnarray}
multiplied by $\frac{36}{k(k+2)}$.
The respective one-point function is equal to
\begin{equation}
\langle G_{-\frac{1}{2}}^{ir}\phi_{n,n}^{ir}
G_{-\frac{1}{2}}^{uv}\phi_{n,n}^{uv}|RG\rangle
=\frac{n^2-5}{n^2-1}\frac{\sqrt{S^{(k-2)}_{1,n}
S^{(k+2)}_{1,n}}}{S^{(k)}_{1,n}}
\end{equation}
}
\end{itemize}
\end{appendix}

\bibliographystyle{JHEP}
\providecommand{\href}[2]{#2}
\begingroup\raggedright

\endgroup

\end{document}